\documentclass[9pt,twocolumn,twoside]{opti_preprint}

\setboolean{shortarticle}{false}

\usepackage{soul}

\title{Temporal recompression through a scattering medium via a broadband transmission matrix}

\author[1,*]{Mickael Mounaix}
\author[2]{Hilton B. de Aguiar}
\author[1]{Sylvain Gigan}

\affil[1]{Laboratoire Kastler Brossel, ENS-PSL Research University, CNRS, UPMC Sorbonne Universit\'{e}s, Coll\`{e}ge de France, 24 rue Lhomond, 75005 Paris, France }
\affil[2]{D\'{e}partement de Physique, Ecole Normale Sup\'{e}rieure / PSL Research University, CNRS, 24 rue Lhomond, 75005 Paris, France}

\affil[*]{Corresponding author: mickael.mounaix@lkb.ens.fr}

\begin{abstract}
The transmission matrix is a unique tool to control light through a scattering medium. A monochromatic transmission matrix does not allow temporal control of broadband light. Conversely, measuring multiple transmission matrices with spectral resolution allows fine temporal control when a pulse is temporally broadened upon multiple scattering, but requires very long measurement time. Here, we show that a single linear operator, measured for a broadband pulse with a co-propagating reference, naturally allows for spatial focusing, and interestingly generates a two-fold temporal recompression at the focus, compared with the natural temporal broadening. This is particularly relevant for non-linear imaging techniques in biological tissues.
\end{abstract}

\begin{document}

\maketitle
\pagestyle{plain}
\ifthenelse{\boolean{shortarticle}}{\abscontent}{}

When monochromatic coherent light propagates in a medium with high refractive index inhomogeneities, it quickly develops into a speckle. Despite the complex structure of speckle patterns, each speckle grain has a deterministic relation to the input fields~\cite{rotter_light_2017}. Over the last decade, wavefront shaping has turned to be an efficient tool to control monochromatic light through highly scattering systems~\cite{Vellekoop2007}, notably by exploiting the transmission matrix~\cite{popoff_measuring_2010}.

Under illumination with a source of large bandwidth, each spectral component can generate a different speckle pattern~\cite{mosk_controlling_2012}. 
Therefore one needs to adjust these additional spectral/temporal degrees of freedom to temporally control the output pulse~\cite{andreoli_deterministic_2015}.
This can be achieved using methods such as nonlinear optical processes~\cite{katz_focusing_2011,aulbach2012spatiotemporal}, time-gating~\cite{aulbach_control_2011,mounaix_deterministic_2016}, and frequency-resolved measurements~\cite{mccabe_spatio-temporal_2011,mounaix_spatiotemporal_2016}. However, these methods underly either low signal-to-noise measurements (non-linear processes), or stability issues as they require lengthy acquisition procedures~\cite{carpenter_complete_2016} and the need of external reference.

An alternative approach is to use self-referencing signals, at the expense of lacking control on spectral degrees of freedom. Recently, "broadband wavefront shaping" experiments reported outcomes disparate from what is expected from monochromatic wavefront shaping, such as a decrease in the independent spectral degrees of freedom~\cite{paudel_focusing_2013,Hsu2015}, and recovery of pure polarization states~\cite{de_aguiar_polarization-resolved_2015}. Notably, these results could have an impact on biomedical imaging~\cite{de_aguiar_enhanced_2016}. Nonetheless, the exact temporal properties of the obtained output pulse via broadband wavefront shaping remain elusive. 
In this letter, we report the first characterization of the so-called broadband transmission matrix of a scattering medium. We exploit it for focusing purposes, and we analyze and interpret its temporal behavior. Unexpectedly, the characterized average pulse length is shorter than the pulse propagating without shaping.

\begin{figure}[!t]
\centering
\includegraphics[scale=0.75]{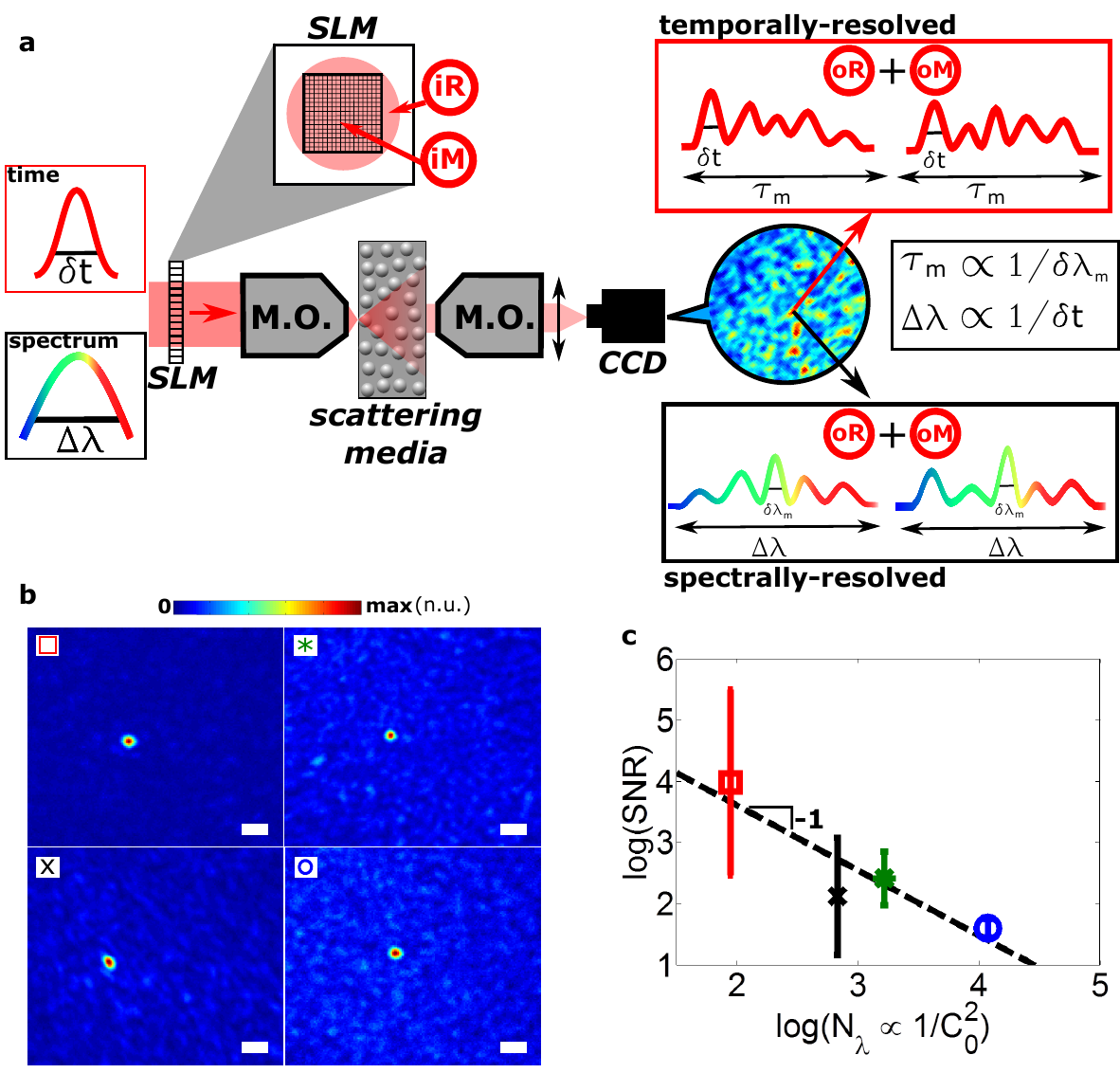}
\caption{\label{schema_speckle}Principle of the measurement of the Broadband Transmission Matrix (BBTM). (a) An input pulse of duration $\delta t$ (spectral width $\Delta \lambda$)  illuminates a spatial light modulator (SLM) before propagation through a thick scattering medium via a microscope objective (MO). The active part of the SLM will modulate a subpart of the input beam ${iM}$ with respect to a static reference part ${iR}$. The transmitted light results in an intensity speckle with low contrast $C_0$ on a CCD. Every single grain is the resulting broadband interference of the transmitted speckle from the input reference part $oR$ and from the input modulated part $oM$. The BBTM is the linear relation connecting $iM$ to $oM$. In the temporal domain, the speckle grain is characterized by its temporal feature $\delta t$, and by the averaged traversal time of photons through the medium $\tau_m$. In the spectral domain, the speckle has a characteristic width $\delta \lambda_m$ which is the spectral speckle correlation bandwidth of the medium. (b) Phase-conjugation of the BBTM enables spatial focusing in a given speckle grain, whose efficiency is inversely related to $N_\lambda$. The images show the results for 4 different samples whose spectral degrees of freedom $N_\lambda \propto 1/C_0^2$ are indicated with corresponding markers in (c). Scale bar: 2$\mu m$. (c) Log of signal-to-noise-ratio (SNR) is plotted as a function of log($N_\lambda$). Error bars are standard deviation for SNR over 50 different focus. Dashed line: linear fit, with a -1 slope.}
\end{figure}

Figure~\ref{schema_speckle}a illustrates and summarizes propagation of broadband light (ultrashort pulse of duration $\delta t$, spectral width $\Delta \lambda$) through an optically thick scattering medium. Transmitted light results in a speckle intensity pattern with low contrast $C_0 < 1$. This low contrast results from the incoherent summation of various uncorrelated speckles corresponding to different spectral components of the input pulse~\cite{Curry2011}. 
The speckle spectral correlation of the medium is characterized by its spectral bandwidth $\delta \lambda_m$. Therefore, the number of independent spectral channels for the pulse, or number of spectral degrees of freedom, is defined as $N_\lambda = \Delta \lambda/ \delta \lambda_m \simeq 1/C_0^2$~\cite{Curry2011,andreoli_deterministic_2015}. In the temporal domain, correspondingly, a speckle grain is characterized by a temporally lengthened structure, with a temporal correlation $\delta t$, related to the bandwidth of the source~\cite{Shi2013,mounaix_deterministic_2016}.  Averaging over different speckle grains leads to retrieving the distribution of optical path lengths of transmitted light through the scattering medium, i.e. the time-of-flight distribution, with a width corresponding to the average traversal time of the photons~$\tau_m \propto 1/ \delta \lambda_m$ \cite{mounaix_spatiotemporal_2016}.

In order to measure a transmission matrix, one needs to measure the output field amplitude, and a convenient methods consists in using phase-stepping interferometry of the speckle with a reference field~\cite{popoff_measuring_2010}. The method we present is a similar to~\cite{popoff_measuring_2010} except that the source is broadband. Figure~\ref{schema_speckle}a illustrates the principle to measure the Broadband Transmission Matrix using a co-propagative ("internal") reference field.
The phase-only SLM is divided into a reference zone and an active zone, which is modulated with respect to the reference part~\cite{popoff_measuring_2010}. 

For a monochromatic input beam at wavelength $\lambda_0$, the measured output field at a pixel $x$, via holographic methods, can be directly related to the input field on the SLM plane at a pixel $y$  with a linear relationship, that is the monochromatic transmission matrix (TM) $T_{\lambda_0}(x,y)$. The experimental measurement of the TM with a co-propagative reference leads to an effective transmission matrix $T'_{\lambda_0}= S_{ref}^* T_{\lambda_0}$, where $S_{ref}$ is a diagonal matrix related to the contribution of the reference field in amplitude and phase~\cite{popoff_measuring_2010} and $^*$ represents the phase-conjugation operation. 

This monochromatic approach breaks down under the illumination of broadband light, such as an ultrashort pulse of spectral-width $\Delta \lambda$. The TM approach has been extended in the spectral domain with the spectrally-resolved multispectral transmission matrix (MSTM) $T(x,y,\lambda)$, which contains $N_\lambda$ monochromatic TMs spectrally separated by $\delta \lambda_m$. Regardless of the reference beam being external~\cite{mounaix_spatiotemporal_2016} or co-propagative~\cite{andreoli_deterministic_2015}, spectral diversity of the medium can be exploited for spectral control. Nevertheless, measurement of the MSTM is a long process as it requires to individually measure each spectral component of the pulse. 

\begin{figure*} [!t]
\centering
\includegraphics[scale=1]{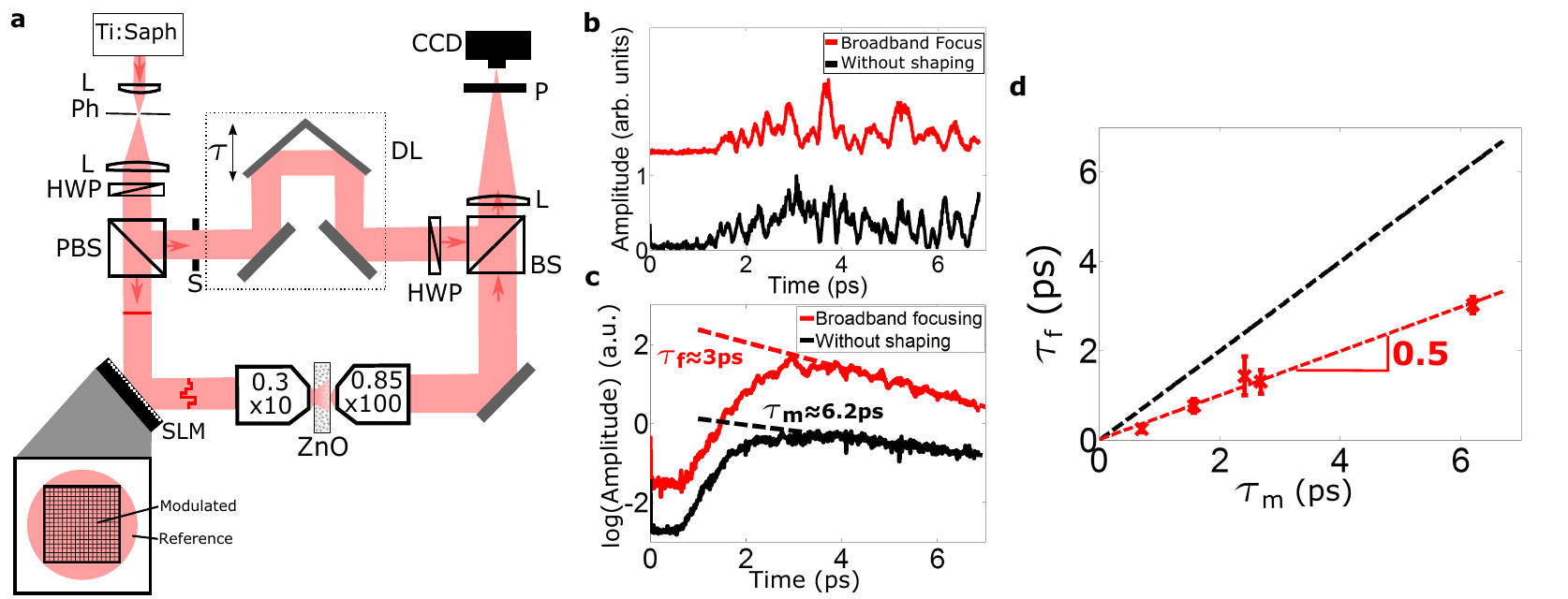}
\caption{\label{short_time} Broadband focusing leads to a shortened averaged confinement time of focused photons. (a) Apparatus for measuring the BBTM and temporal/spectral characterization of the achieved focus. An ultra-short pulse generated by a Ti:Sapphire laser ($\lambda_0=$ 800~nm, $\tau_0 \approx 100$ fs FWHM, MaiTai, Spectra-Physics)  is reflected by a phase-only SLM (X10468-2, Hamamatsu) and injected into a thick scattering sample of randomly distributed ZnO nanobeads using a Microscope Objective (MO). Transmitted scattered light is collected with another MO, and imaged on a CCD after a polarizer (P). A reference arm is added, by selecting part of the laser light before the SLM on a polarizing beam splitter and recombining with the output scattered light on a beamsplitter (BS) before the CCD camera. A delay line (DL)  controls the relative optical path delay between the reference and the scattered pulse. Importantly, this arm is blocked by a shutter (S) during both BBTM measurement and focusing experiment, and is only used for Interferometric Cross correlation (ICC) of the output speckle. (L): Lens; (Ph): Pinhole. (b) Time-of-flight distributions retrieved with ICC, in a single speckle grain in the case of (red) focusing with the BBTM and (black) without shaping. The two plots are shifted for better visibility. (c) Corresponding averaged time-of-flight distributions over 25 different foci plotted in semilog. $\tau_{m}$ (without shaping) and $\tau_{f}$ (broadband focusing) are measured with a linear fit. (d) Comparison between $\tau_{m}$ and $\tau_{f}$ for various samples. While dashed black line represents the expected $\tau_{m} = \tau_{f}$, experimental data (red markers) evidences $\tau_{f}$ is twice smaller than $\tau_{m}$. Red dashed line represents a linear fit and error bars indicate the standard deviation for estimation of $\tau_{f}$.}
\end{figure*}

We now assume the input source is broadband, and the reference beam is co-propagative as illustrated in Figure~\ref{schema_speckle}. Transmitted light through the scattering medium, measured with a CCD camera without spectral resolution, results in a low-contrasted intensity speckle. Indeed, this speckle can be seen, for the sake of simplification, as the incoherent superposition of $N_\lambda$ uncorrelated monochromatic speckles. Each monochromatic speckle can be decomposed as a coherent sum between a reference field $E^{out,R}$ and a modulated field $E^{out,M}$ over a spectral interval $\delta \lambda_m$. 

The transmitted intensity can be written as a static term $I_0$, which contains all field-squared terms, and the incoherent sum of $N_\lambda$ uncorrelated cross terms between reference and modulated fields. In essence, each individual cross term is related to its corresponding monochromatic TM coefficients at a given wavelength $\lambda_i$: $E^{out,M}(\lambda_i)=T(\lambda_i) E^{in}(\lambda_i)$, where $E^{in}(\lambda_i)$ is the controlled input field at the SLM pixel (see iM in Fig. \ref{schema_speckle}a).Thus, the transmitted intensity reads:

\begin{equation} \label{intensity_BBTM}
\begin{split}
I & = \sum_{\lambda_i=1}^{N_\lambda}I(\lambda_i)= \sum_{\lambda_i=1}^{N_\lambda} \vert E^{out,R}(\lambda_i) + E^{out,M}(\lambda_i) \vert ^2 \\
& = I_0 + \sum_{\lambda_i=1}^{N_\lambda} \bigg[ \Big(E^{out,R}(\lambda_i)\Big)^* E^{out,M}(\lambda_i) \quad \mbox{+ c.c.} \bigg] \\
& =  I_0 + \sum_{\lambda_i=1}^{N_\lambda} \bigg[ \underbrace{\Big(E^{out,R}(\lambda_i)\Big)^* T(\lambda_i)}_{T'(\lambda_i)} E^{in}(\lambda_i) \bigg]
\end{split}
\end{equation}

A retardation is now applied on the modulated part of the broadband input field. As long as the pulse can be considered narrowband, meaningè $\Delta \lambda / \lambda_0 \ll 1$, this retardation translates into a phase shift $\varphi$ that is identical for all the spectral components of the input pulse: $\forall \lambda_i \in \Delta \lambda, \varphi(\lambda_i)=\varphi$. Consequently, the $N_\lambda$ spectral modes will experience the same phase shift $\varphi$.
Therefore, modulating the phase of a pixel of the SLM implies the modulation of the output broadband intensity speckle. A complex, although measurable, linear operator is then connecting the phase of every controllable pixel on the SLM and the output intensity. We name this operator the Broadband Transmission Matrix $B$ (BBTM), which as evident from Eq.~\ref{intensity_BBTM} is the incoherent sum of $N_\lambda$ monochromatic effective TMs: $B = \sum_{\lambda_i=1}^{N_\lambda} T'(\lambda_i)$.

Experimentally, performing the same measurement as for the monochromatic case~\cite{popoff_measuring_2010}, but for a pulse, gives access to this broadband transmission matrix $B$. $B$ is measured using Hadamard basis of SLM pixels in order to benefit from higher signal-to-noise~\cite{popoff_measuring_2010}. An \textit{a posteriori} change of basis enables to get the BBTM in the canonical basis of the SLM pixels.

We first quantify the conventional focusing properties of the BBTM in the spatial domain. Figure \ref{schema_speckle}b shows representative results for 4 samples of ZnO powder of different thicknesses. 
Similarly to the monochromatic approach, phase-conjugating a line of the BBTM enables focusing the output pulse in a chosen speckle grain. Signal-to-noise ratio (SNR) of the achieved broadband focus is defined as the ratio between $I^{\text{focus}}$ intensity at the focus over $\langle I^{\text{random}} \rangle $, the averaged background speckle intensity. 
Using a broadband source, the values of SNR is expected to scale as $\propto N_{SLM}/N_\lambda$~\cite{lemoult_manipulating_2009}. The dependence of SNR regarding $N_\lambda$ is experimentally verified in Figure \ref{schema_speckle}c.

To go further, after acquiring $B$ and focusing, the temporal profile of the output pulse in the focus can be measured with a linear Interferometric Cross-Correlation technique~\cite{mounaix_spatiotemporal_2016}. The shutter S of Fig.~\ref{short_time}a is now opened to allow the external reference pulse to sample either the focus or a random speckle grain. Briefly, the method consists in recording the interferogram between the input ultrashort plane wave and the output stretched speckle on th CCD camera.

Figure \ref{short_time}b shows representative time-domain results for a medium with $N_\lambda \simeq$ 60. Examples of temporal profiles of a single grain are presented without wavefront shaping (black line) and at the focus (red line).  
Averaging over many spatial grains enables to retrieve the time-of-flight distribution of the medium, and consequently $\tau_m$. Identically, averaging temporal profiles of foci over different realizations enables to retrieve $\tau_{f}$, the average temporal duration at the focus. For the given thick scattering sample, we obtain $\tau_{m} = 6.2$ ps in Fig.~\ref{short_time}c. Remarkably, with the very same sample we obtain a shorter time $\tau_{f}  \simeq 3$ ps $< \tau_{m}$. Therefore, the focus is temporally shorter than the medium.  Fig.~\ref{short_time}d shows the observed values of $\tau_{f}$ and $\tau_{m}$ upon varying optical thickness.
The dashed black line represents the expected $\tau_{m} = \tau_{f} $, from a spatial-only focusing experiment. Clearly, we observe that $\tau_{f}$ is systematically smaller than $\tau_{m}$, precisely, $\tau_{f} /\tau_{m} =0.5$: this is the main result of the paper.

This ratio of 2 can be easily understood: the BBTM coefficients are inherently cross-terms between reference and modulated broadband speckles as written in Equation~\ref{intensity_BBTM}, that have both similar time-of-flight distributions with, on average, a decay of the form $\propto e^{-t/\tau_m}$. Therefore this cross-term, has then a decay of the form $\propto e^{-2t/\tau_m}$ on average. The corresponding temporal duration at the focus $\tau_{f}\simeq \tau_{m} / 2$ is consequently two times lower. We further corroborate these results with spectral speckle autocorrelation. A decrease of $\tau_{f}$ leads to a spectral correlation bandwidth $\delta \lambda_{f}$ larger than $\delta \lambda_{m}$ (not shown).

\begin{figure} [!t]
\centering
\includegraphics[scale=0.75]{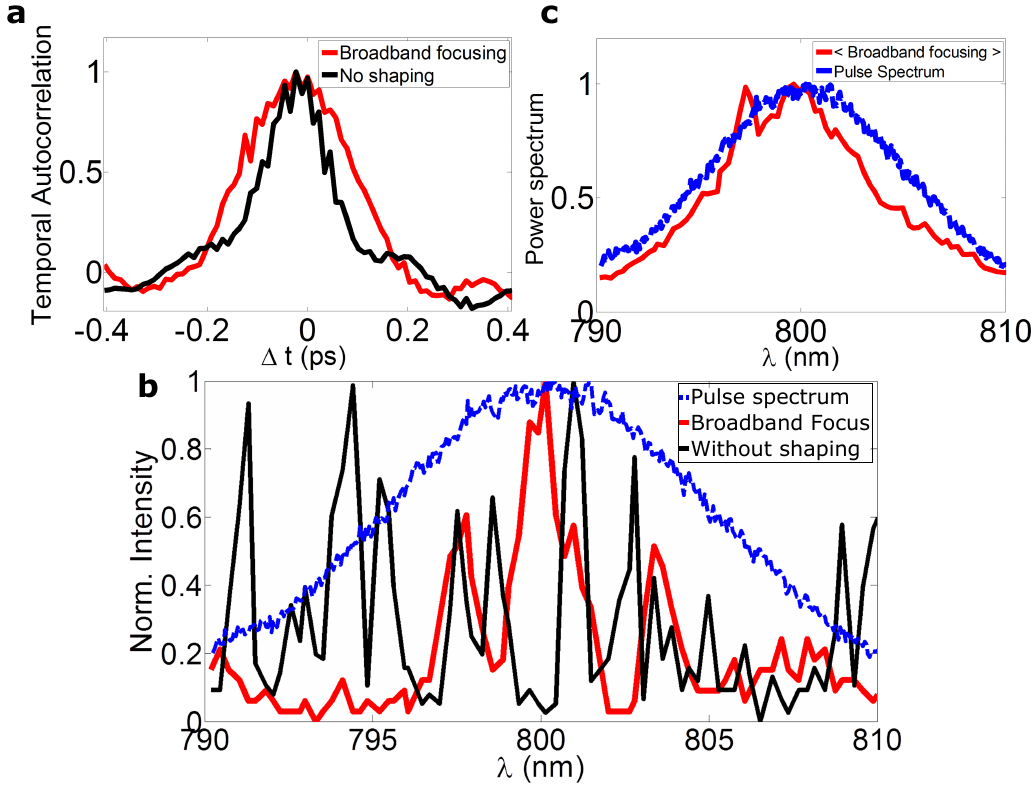}
\caption{\label{narrow_spectrum} Broadband focusing changes the envelop spectral response.  (a) Temporal field autocorrelation of both (red) focus averaged over 25 different foci and (black) without shaping averaged over 200 speckle grains. The temporal width of the autocorrelation peak for the focus is broader than without shaping. (b) Spectral intensity measurement. Monochromatic intensity in a single grain (with or without shaping) is recorded as function of $\lambda$. Plots are spectral intensities (black) without wavefront shaping, (red) at the focus, and (blue) intensity spectrum of the input pulse. (c) (red) Averaged spectral intensity of focus over 100 different focus compared to (blue) the spectrum of the input pulse. On average, the output pulse has a narrower spectrum than the input pulse.}
\end{figure}

The results above analyze averaged properties of temporal profiles, which are related to the smallest features in the frequency-domain. Another relevant, yet independent, information is located in the fast variation of the temporal fluctuations, which is the temporal speckle grain. To extract this quantity from Fig.~\ref{short_time}b, temporal autocorrelations are plotted in Fig.~\ref{narrow_spectrum}a for both (red) a focus and (black) a grain corresponding to a random input pattern. It is clearly seen that $\delta t_{f} >  \delta t_{m}$. As $\delta t \propto 1/\Delta \lambda$, a direct consequence is the spectral narrowing of the bandwidth. 

We now focus on the averaged frequency-domain response. For spectral characterization of the focus, the laser is unlocked, and it is used as a tunable cw source~\cite{andreoli_deterministic_2015}. For every single wavelength, a CCD image is recorded either when focusing or for a random input mask~\cite{van2011frequency} with a scattering medium with $N_\lambda \simeq$ 20. In Fig.~\ref{narrow_spectrum}b, we present spectral intensity (black line) of a single speckle grain for a random input phase pattern. As expected, around $N_\lambda$ spectral grains can be seen within the pulse spectral width. Conversely, for the focus (red line), fewer grains are observed, and are mostly contributing to the central frequency. The average over 100 different foci (red, Fig.~\ref{narrow_spectrum}c), shows a narrower spectrum than the Gaussian spectrum of the input pulse (blue). 

Accordingly, this observation can be explained using Equation~\ref{intensity_BBTM}. Both reference and modulated speckles have the same spectral profile than the input pulse, which is a Gaussian centered at $\lambda_0$ over a span $\Delta \lambda_{m}$. Therefore, the cross product is a narrower Gaussian with $\Delta \lambda_{f} = \Delta \lambda_{m}/\sqrt{2}$ at FWHM, which is consistent with experimental data.

In conclusion, we have characterized the temporal focus obtained via the Broadband Transmission Matrix measured with co-propagative reference. BBTM can be straightforwardly measured, and it leads to surprising temporal and spectral properties of the output pulse. Indeed, with a Gaussian input pulse, we have shown that focusing with BBTM enables to reduce $N_\lambda$ by a factor $\sim 2 \sqrt{2}$, which comes from a factor 2 in the increase of spectral correlation bandwidth, and a factor $\sqrt{2}$ in the shortening of the spectral envelope. This result may have practical implications in nonlinear microscopy for deep imaging, as in practice it allows relaxing the need to control the spectral degrees of freedom at few mm-thickness specimens~\cite{katz_focusing_2011}.

\paragraph{Funding Information}
This work was funded by the European Research Council (COMEDIA Grant No. 278025 and SMARTIES Grant No. 724473). H.B.A is supported by LabEX ENS-ICFP: ANR-10-LABX-0010/ANR-10-IDEX-0001-02 PSL. S.G. is a member of the Institut Universitaire de France.  

\paragraph{Acknowledgments}

The authors thank Sophie Brasselet for related discussions.

\bibliographystyle{apsrev4-1}
\bibliography{biblio_broadband}

\end{document}